\documentclass[prc,aps,12pt,final,notitlepage,oneside,onecolumn,
nobibnotes,nofootinbib,superscriptaddress,noshowpacs]{revtex4-1}
\usepackage{graphicx,colordvi}
\usepackage{amssymb,amsmath}
\usepackage{array}
\usepackage{calc}
\usepackage{ifthen}
\usepackage{epsfig}

\def\d{\hbox{d}}
\def\be{\begin{equation}}
\def\ee{\end{equation}}
\def\bea{\begin{eqnarray}}
\def\eea{\end{eqnarray}}
\def\l{\label}
\def\r{{\bf r}}

\def\hahat{\hat{H}}

\def\hahat0{\hat{H}_0}

\def\bs{\bigskip}
\def\ms{\medskip}

\def\cos{\hbox{cos}}
\def\sin{\hbox{sin}}

\def\d{\hbox{d}}
\def\eps{\epsilon}
\def\epsd{\varepsilon}

\def\epsi{{\cal E}}
\def\siml{\hbox{\kern.1em \lower.6ex \hbox{$\sim$} \kern-1.12em
 \raise.6ex \hbox{$<$} \kern.1em}}
\def\simg{\hbox{\kern.1em \lower.6ex \hbox{$\sim$} \kern-1.12em
 \raise.6ex \hbox{$>$} \kern.1em}}

\begin{document}

\title{NUCLEAR ASYMMETRY ENERGY AND ISOVECTOR STIFFNESS
WITHIN THE EFFECTIVE SURFACE APPROXIMATION}
\author{J.P. Blocki}
\affiliation{\it National Centre for Nuclear Research,
Otwock 05-400, Poland}
\author{A.G. Magner\footnote{magner@kinr.kiev.ua}}
\affiliation{\it Institute for Nuclear Research,  Kyiv 03680, Ukraine}
\author{P. Ring}
\affiliation{\it Technical Munich University,  D-85747 Garching, Germany}
\author{A.A. Vlasenko}
\affiliation{\it National Technical University of Ukraine "KPI", Kyiv 03056, Ukraine}

\vspace*{-1.05cm}

\begin{abstract}
The isoscalar and isovector particle densities in the effective surface 
approximation to
the average binding energy are used to derive analytical expressions of the 
surface symmetry energy, the neutron skin thickness
and the isovector stiffness of sharp edged proton-neutron asymmetric nuclei.
For most Skyrme forces the isovector coefficients of the surface energy 
and of the stiffness are significantly different from the 
empirical values derived in the liquid drop model.
Using the analytical isovector surface energy constants 
in the framework of the hydrodynamical
and the Fermi-liquid droplet models the mean energies and the 
sum rules of the isovector giant
dipole resonances are found to be 
in fair agreement with the experimental data.

\end{abstract}
\maketitle
\date{\today }

 \section{Introduction}

 A simple and accurate solution of particle-density distributions was 
obtained within the nuclear effective surface (ES)
approximation in Refs.\ \cite{strtyap,strmagbr,strmagden}.
It exploits the saturation properties of nuclear matter
in the narrow diffuse-edge region in finite heavy nuclei.
The ES is defined as the location of points with a maximum density gradient.
An orthogonal coordinate system related locally to the ES is specified by
the distance $\xi$ of a given point from this surface and tangent coordinates
$\eta$ parallel to the ES (see Fig.\ \ref{fig1}). Using nuclear energy density 
functional
theory, the variational condition derived from minimizing the nuclear energy 
at some fixed
integrals of motion is simplified in the $\xi,\eta$ coordinates. 
In particular, in the extended Thomas-Fermi
(ETF) approach \cite{brguehak}, it can be done for any fixed deformation using 
the expansion
in a small parameter $a/R \sim A^{-1/3} \ll 1$ for heavy enough nuclei, 
where $a$ is of the order of
the diffuse-edge thickness of the nucleus, 
$R$ is the mean curvature radius of the ES, and
$A$ is the number of nucleons. The accuracy of the ES  approximation 
in the ETF approach
without spin-orbit (SO) and asymmetry terms was checked \cite{strmagden}
by comparing results with those of the Hartree-Fock (HF) and ETF theories
for some Skyrme forces. The ES approach \cite{strmagden} was also extended by
taking into account the SO and asymmetry effects \cite{magsangzh}.

In the present work, solutions for the isoscalar and isovector 
particle densities and energies in the ES
approximation of the ETF approach are applied to analytical 
calculations of the surface symmetry energy, the neutron skin, and the isovector
stiffness coefficient in the leading order of the parameter $a/R$ 
(see also Ref. \cite{BMV}). Our results are
compared with older investigations 
\cite{myswann69,myswnp80pr96,myswprc77,myswiat85} in the liquid droplet model
(LDM) and with more recent works 
\cite{danielewicz1,pearson,danielewicz2,vinas1,vinas2,vinas3,vinas4,kievPygmy,nester}.
We suggest also studying also the splitting of the isovector giant dipole 
resonances into main and satellite (pygmy) peaks
\cite{kievPygmy,nester}  as a function of the analytical isovector surface 
energy constant of the ES approach
within the  Fermi-liquid droplet (FLD) model \cite{denisov,kolmag,kolmagsh}. 
The analytical expressions for the surface symmetry
energy constants are tested by the mean energies of the isovector giant 
dipole resonances (IVGDR) within the
hydrodynamical (HD) and FLD models.

The manuscript is organized as follows:
In Sec. II we give an outlook of the basic points of the ES approximation
within the density functional theory, and the main results 
for the isoscalar and 
isovector
particle densities. Section III is devoted to analytical derivations of the
symmetry energy in terms of the surface energy coefficient, the neutron skin 
thickness,
and the isovector stiffness. The discussions of the results are given in
Sec. IV and summarized in Sec. V.
Some details of our calculations are presented in Appendixes A-C.

\section{Energy and particle densities}

We start with the nuclear energy as a functional of the isoscalar and the isovector densities
$\rho_{\pm}=\rho_n \pm \rho_p$:
\be\l{energy}
E=\int \d \r\; \epsi\left(\rho_{+},\rho_{-}\right)\;,
\ee
in the local density approach
\cite{brguehak,chaban,reinhard,bender,revstonerein,ehnazarrrein}
with the energy density $\epsi\left(\rho_{+},\rho_{-}\right)$,
\bea
\epsi\left(\rho_{+},\rho_{-}\right) &\approx&
- b^{}_V \rho_{+}
+ J I^2 \rho_{+} + \rho_{+} \left[\epsd_{+}(\rho_{+}) -
\epsd_{-}(\rho_{+},\rho_{-})\right]+
\nonumber\\
&+&
\left({\cal C}_{+} +{\cal D}_{+} \rho_{+} \right)
\left(\nabla \rho_{+}\right)^2
+ \left({\cal C}_{-} + {\cal D}_{-} \rho_{+}\right)
\left(\nabla \rho_{-}\right)^2\;,
\label{enerden}
\eea
where $~I=(N-Z)/A$ is the asymmetry parameter,
$~N=\int \d \r \rho_n(\r)$ and $~Z=\int \d \r \rho_p(\r)$ are
the neutron and proton numbers and $~A=N+Z$. As usual, the energy density
${\cal E}$ in Eq.\ (\ref{enerden}) contains
the volume part given by the first two terms of Eq.\ (\ref{enerden})
and the surface part including the density gradients
\cite{strtyap,strmagden}. The particle separation energy
$b^{}_V \approx$ 16 MeV
and  the symmetry energy constant of the nuclear matter $~J \approx$
30 MeV specify the volume terms in Eq.\ (\ref{enerden}).  
Equation (\ref{enerden})
can be applied in a semiclassical approximation for realistic
Skyrme forces \cite{chaban,reinhard,bender,revstonerein,ehnazarrrein},
in particular by neglecting higher $\hbar$ corrections
in the ETF kinetic energy \cite{brguehak,strmagbr,strmagden}
and Coulomb terms. Up to small Coulomb exchange terms they all can be easily
taken into account (see Refs. \cite{strtyap,strmagden,magsangzh}).
The constants ${\cal C}_{\pm}$ and ${\cal D}_{\pm}$ are defined by the
parameters of the Skyrme forces~\cite{chaban,reinhard}. The isoscalar part
of the surface energy density, which does not depend explicitly
on the density gradient terms, is determined by the function
$\epsd_{+}(\rho_{+})$~\cite{strmagden,magsangzh},
which satisfies the saturation condition
$\epsd_{+}(\overline{\rho})=0$, 
$(\d\epsd_{+}(\rho_{+})/\d\rho_{+})_{\rho_{+}=\overline{\rho}} =0$,
where $\overline{\rho}=3/4\pi r_0^3 \approx$ 0.16 fm$^{-3}$
is the density of the infinite nuclear matter and $r_0 = R/ A^{1/3}$ 
is the radius constant.
Here we use a quadratic approximation, 
$\epsd_{+}=K (\rho_{+}-\overline{\rho})^2/(18 \overline{\rho}^2)$,
where $K$ is the incompressibility modulus of symmetric nuclear matter, mainly $K \approx 220-260$ MeV (see Table I).
The isovector component can be simply evaluated as 
$\epsd_{-}=J\;\left(I^2-\rho_{-}^2/\rho_{+}^2\right)$~\cite{magsangzh}.
The isoscalar SO gradient terms in Eq.\ (\ref{enerden}) are defined 
with a constant:
${\cal D}_{+} = -9m W_0^2/16 \hbar^2$, where
$W_0 \approx$100 - 130~ MeV$\cdot$fm$^{5}$ and
$m$ is the nucleon mass~\cite{brguehak,chaban,reinhard,bender,revstonerein}.

Minimizing the energy $E$ under the constraints
of the fixed particle number $A=\int \d \r\; \rho_{+}(\r)$ and
 neutron excess $N-Z= \int \d \r\; \rho_{-}(\r)$
(also others, such as deformation
 \cite{strtyap,strmagden}),
one arrives at the Lagrange equations with the corresponding multipliers,
$\lambda_{+}$ and $\lambda_{-}$ being the
isoscalar and isovector chemical potentials, 
respectively (see Appendixes A and B).
Our approach can be applied for any deformation parameter of the nuclear
surface if its diffuseness with respect to the curvature radius
$a/R$ is small. The analytical solutions will be obtained approximately 
up to the order of $A^{2/3}$ in the
binding energy. To satisfy the condition of particle number conservation
with the required accuracy, we account for relatively small
surface corrections ($\propto a/R \sim A^{-1/3}$ at the first order)
to the leading terms in the Lagrange multipliers
\cite{strmagbr,strmagden,magsangzh}
(Appendixes B and C).

For the isoscalar particle density, $w =\rho_{+}/\overline{\rho}$ one has
up to {\it leading} terms in the parameter  $a/R$ the usual first-order
differential Lagrange equation with the solution \cite{strmagden,magsangzh}
\be\l{ysolplus}
x=-\int_{w_r}^{w}\d y\; \sqrt{\frac{1 +\beta y}{y\eps(y)}}\;,\qquad
x=\frac{\xi}{a}\qquad
\left(a=\sqrt{\frac{{\cal C}_{+}\; \overline{\rho}\; K}{30\; b_V^2}}\right)
\ee
below the turning point $x(w=0)$; $w=0$ for $x \geq x(w=0)$ and
$\beta={\cal D}_{+}\overline{\rho}/{\cal C}_{+}$ is the dimensionless SO
parameter. For convenience we introduced also the dimensionless parameter
 $\eps=18\epsd_{+}/K$.  For $w_r=w(x=0)$
one has the boundary condition
$\d^2 w(x)/\d x^2=0$
at the ES ($x=0$):
\begin{equation}
\eps(w_r)+w_r(1 +\beta w_r) \left[\d \eps(w)/\d w\right]_{w=w_r}=0\;\;.
\label{boundcond}
\end{equation}
In Eq.\ (\ref{ysolplus}), $a \approx 0.5-0.6$ fm is the
diffuseness parameter as shown in Table I ($\xi = r-R~$ for
spherical nuclei in spherical coordinates).
The diffuseness of the edge $a^{}_d$
is given by
\be\l{ad}
a^{}_d=\sqrt{5(\overline{\xi^2}-\overline{\xi}^2)/3} =
a \sqrt{5(\overline{x^2}-\overline{x}^2)/3}\;,\qquad
\overline{x^n}=\int_{-\infty}^{\infty} x^n \d x (\d w/\d x)\;,
\ee
where bars mean an averaging with the surface density distribution
$dw/dx$ \cite{strmagden}.
For all Skyrme forces (Table I) the parameter $a$  introduced in
Eq.\ (\ref{ysolplus}) measures the diffuseness of the nuclear edge as
the mean-squared fluctuation of
$\xi$ due to the relation $a_d \approx a \sqrt{5/3}$
(see also Refs.\ \cite{strmagden,vinas2}).
As shown in Ref. \cite{magsangzh},  the influence of the
 semiclassical $\hbar$ corrections (related to the ETF
kinetic energy) to $w(x)$ is negligibly small everywhere,
besides the quantum tail outside
the nucleus ($x \simg 1$). Therefore, all these corrections were neglected in
Eq.\ (\ref{enerden}).
With a good convergence of the expansion of the
$\eps(y)$ in powers of $1-y$
 up to the quadratic term \cite{strmagden,magsangzh},
$\eps=(1-y)^2$,
one finds the analytical solutions of Eq.\ (\ref{ysolplus}) in terms of 
algebraic, trigonometric, and
logarithmic functions [see Eq.\ (\ref{ysolan})].
For $\beta=0$ (i.e., without SO terms), it simplifies to
the solution $w(x)=\tanh^2\left[(x-x_0)/2\right]$ for
$x\leq x_0=2{\rm arctanh}(1/\sqrt{3})$
and zero for $x$ outside the nucleus ($x>x_0$).

For the isovector density, $w_{-}(x)=\rho_{-}/(\overline{\rho}I) $,
after simple transformations of the isovector Lagrange equation
up to the leading term in $a/R$  in the ES approximation one similarly
finds the
equation and boundary condition [see Eq.\ (\ref{yeq0minus})].
The analytical solution for $w_{-}=w \cos[\psi(w)/\sqrt{1+\beta}]$
can be obtained through the expansion (\ref{powser}) of $\psi$ in powers of
\be\l{csymwt}
\widetilde{w}(w)=(1-w)/c_{sym}\qquad \mbox{ with} \qquad
c_{sym}=a \;\sqrt{\frac{J}{\overline{\rho}\;
\vert{\cal C}_{-}\vert}}\;.
\ee
Expanding up to the second order in $\widetilde{w}$ one finds (Appendix A)
\be\l{ysolminus}
w_{-} =  w\;\left(1-\frac{\psi^2(w)}{2\left(1+\beta\right)}
\right),\qquad
\psi(w)=\widetilde{w}(w) \left[1
+ \tilde{c}\widetilde{w}(w)\right],\qquad
\widetilde{c}=\frac{\beta c_{sym}/2-1}{1+\beta}\;.
\ee
In Fig.\ \ref{fig2} the SO dependence of the function $w_{-}(x)$ is compared with
that of the density $w(x)$ for the SLy7 force as a typical example 
\cite{magsangzh}.
It might seem from a brief look at Fig.\ \ref{fig3} that  
the isovector $w_{-}(x)$ [and therefore, the isoscalar
$w(x)$] densities depend weakly on the
most of the Skyrme forces \cite{chaban,reinhard}. However, as  shown in a 
larger (logarithmic) scale in Fig.\ \ref{fig4},
one observes notable differences in the isovector densities $w_{-}$ derived 
from different Skyrme forces
within the edge diffuseness. In particular, as shown below, this is 
important for the calculations of the neutron skins of nuclei.

We emphasize that the dimensionless densities, $w(x)$ 
[Eqs.\ (\ref{ysolplus}) and (\ref{ysolan}) ] and $w_{-}(x)$ 
[Eq.\ (\ref{ysolminus})], shown in Figs.\ \ref{fig2}-\ref{fig4} were 
obtained in 
the leading ES approximation ($a/R \ll 1$) as functions of the 
specific combinations
of the Skyrme force parameters such as $\beta$ and $ c_{sym}$ of 
Eq.\ (\ref{csymwt}). Therefore, they are the universal distributions
independent of the specific properties of the nucleus such as the neutron and 
proton numbers, and the deformation and curvature of the nuclear ES;
see also Refs.\ \cite{strtyap,strmagden,magsangzh}.
These distributions yield approximately the spatial coordinate dependence
of local densities in the 
direction that is normal to the ES 
with the correct asymptotical behavior outside of the ES layer for any 
ES deformation satisfying the condition $a/R \ll 1$ (in particular,
for the semi-infinite nuclear matter); see further discussions
below.

\section{Isovector energy and stiffness}

Within the improved ES approximation where also
higher order corrections in the small parameter $a/R$ are taken into account we derive equations for the
nuclear surface itself (see Appendix B and Refs.\ \cite{strmagbr,strmagden,magsangzh}).
For more exact isoscalar and isovector particle densities we account for the
main terms in the next order of the parameter $a/R$
in the Lagrange equations [cf.
Eq.\ (\ref{lagrangeqminus1}) as compared with Eq.\ (\ref{lagrangeqminus0})].
Multiplying these equations by $\partial \rho_{-}/\partial \xi$
and integrating them over the ES
in the normal-to-surface direction $\xi$ and using
the solutions for $w_{\pm}(x)$ up to the leading orders
[see Eqs.\ (\ref{ysolplus}) and
(\ref{ysolminus})],
one arrives at the ES equations in the form of
the macroscopic boundary conditions (\ref{macboundcond})
\cite{strtyap,strmagbr,strmagden,magsangzh,kolmagsh,magstr,magboundcond,bormot}.
They ensure equilibrium through the equivalence of the volume and surface 
(capillary) pressure
(isoscalar or isovector) variations.
As shown in Appendix B, the latter ones are proportional to the corresponding
surface tension coefficients:
\be\l{sigma}
\sigma_{\pm}=b_S^{(\pm)}/(4 \pi r_0^2),\qquad b_S^{(\pm)} \approx
8 \pi r_0^2 {\cal C}_{\pm}\int_{-\infty}^{\infty} \d \xi\;
\left(1 + \frac{{\cal D}_{\pm}}{{\cal C}_{\pm}} \rho_{+}\right)
\left(\frac{\partial \rho_{\pm}}{\partial \xi}\right)^2.
\ee

The nuclear energy $E$
[Eq.\ (\ref{energy})]
in this improved ES approximation (Appendix C) is split into volume and
surface (both with the symmetry) terms,
\be\l{EvEs}
E \approx -b^{}_V\; A + J (N-Z)^2/A + E_S\;.
\ee
For the surface energy $E_S$ one obtains
\be\l{Es}
E_S = E_S^{(+)} + E_S^{(-)}
\ee
with the following
isoscalar (+) and isovector (-) surface components:
\be\l{Espm}
 E_S^{(\pm)}= \sigma_{\pm}{\cal S}=b_S^{(\pm)} {\cal S}/(4\pi r_0^2),
\ee
where ${\cal S}$ is the surface area of the ES.
The energies $E_S^{(\pm)}$ in Eq.\ (\ref{Espm}) are determined by the
isoscalar $b_{S}^{(+)}$ and isovector $b_{S}^{(-)}$ surface energy constants of
Eq.\ (\ref{sigma}).
These constants are proportional to the corresponding surface tension coefficients
$\sigma_{\pm}$ in Eq.\ (\ref{sigma}) through the solutions (\ref{ysolplus}) and (\ref{ysolminus})
for $\rho_{\pm}(\xi)$ which can be taken into account in leading order of $a/R$ (Appendix C). These
coefficients $\sigma_{\pm}$ are the same as those found in the
expressions for the capillary pressures of the boundary conditions (\ref{macboundcond}).

For the energy surface coefficients $b_{S}^{(\pm)}$ one obtains
\bea
\l{bsplus}
b_S^{(+)}&=&\frac{6 {\cal C}_{+}
\overline{\rho} {\cal J}_{+}}{ r_0 a},\qquad
{\cal J}_{+}=\int_0^1 \d w
\sqrt{w(1+\beta w) \eps(w)} \;, \\
\l{bsminus}
b_S^{(-)}&=&k^{}_S\; I^2,\qquad~~~~~
k^{}_S= 6 \overline{\rho}\; {\cal C}_{-}\;{\cal J}_{-}/(r_0 a)\;,\\
\vspace{-4ex}
\l{jminus}
{\cal J}_{-}&=&-\frac{1}{1+\beta}\int_0^1 \left(1-w\right)^2 \d w\;
\sqrt{\frac{w(1+\beta w)}{\eps(w)}}\;
\left(1+\widetilde{c}\widetilde{w}\right)^2\;.
\eea
For $\widetilde{w}$ and $\widetilde{c}$, see Eqs.\ (\ref{csymwt}) and (\ref{ysolminus}),
respectively. Simple expressions for the constants
$b_S^{(\pm)}$ in Eqs.\ (\ref{bsplus}) and
(\ref{bsminus}) can be easily derived  in terms of
algebraic and trigonometric functions by calculating explicitly the integrals over $w$
for the quadratic form of $\eps(w)$ [Eqs. \ (\ref{Jp}) and (\ref{Jm})].
Note that in these derivations we neglected curvature terms and, being of the same order,
shell corrections. The isovector energy terms were obtained within the ES
approximation with high accuracy up to the product of two
small quantities, $I^2$ and $(a/R)^2$.

According to the theory \cite{myswann69,myswnp80pr96,myswprc77},
one may define the isovector stiffness $Q$ with respect to the difference
$R_n-R_p$ between the neutron and proton radii as a collective variable,
\be\l{Esq}
E_s^{(-)}=-\frac{\overline{\rho} r_0}{3} \oint
\d S\;Q \tau^2 \approx
-\frac{Q\tau^2 {\cal S}}{4 \pi r_0^2},\qquad  \tau=\frac{R_n-R_p}{r_0}\;,
\ee
where $\tau$ is the neutron skin.
Comparing this expression with Eq.\ (\ref{Espm})
for the isovector surface energy written through
the isovector surface energy constant
$b_S^{(-)}$ [Eq.\ (\ref{bsminus})], one obtains
\be\l{stifminus}
Q=-b_S^{(-)}/\tau^2 =-k^{}_S I^2/\tau^2\;.
\ee
Defining the neutron and proton radii $R_{n,p}$ as the positions of the
maxima of the neutron and proton density gradients, respectively,
\be\l{defrnp}
\left(\frac{\partial^2 \rho_{n,p}}{ \partial r^2}\right)_{r=R_{n,p}}=0,\qquad
\left(\frac{\partial^2 \rho_{+}}{\partial r^2}\right)_{r=R}=0,
\ee
we use the expansion in small values of $\delta R_{n,p}=R_{n,p} -R$ near
the ES. Thus, in the linear approximation in $\delta R_{n,p}$ and $I$ one obtains
\be\l{skin}
\tau=-2\frac{a I}{r_0}\;
\frac{\partial^2 w_{-}}{\partial x^2}\Big|_{x=0}\left(
\frac{\partial^3 w}{\partial x^3}\Big|_{x=0}\right)^{-1}=\frac{8 a I}{
r_0 c_{sym}^2}g(w_r),\,
\ee
where
\be\l{fw}
g(w)=\frac{w^{3/2}(1+\beta w)^{5/2}}{
(1+\beta)(3w+1+4 \beta w)}\;\left\{w
(1+2\widetilde{c} \widetilde{w})^2
+2\widetilde{w}  \left(1+\widetilde{c}
\widetilde{w}\right) \left[\widetilde{c}
w - c_{sym}\left(1+
2 \widetilde{c} \widetilde{w}\right)\right]\right\},
\ee
and $w_r$ is the solution of the boundary equation (\ref{boundcond}).
In the derivations of Eq.\ (\ref{skin}), we used the approximation
$\eps(w)=(1-w)^2$ and
expressions (\ref{ysolplus}) for $w(x)$ and (\ref{ysolminus}) for
$w_{-}(x)/w$.
The neutron and proton particle-density variations in
Eq.\ (\ref{defrnp}) conserve the center of mass in the same
linear approximation
in  $\delta R_{n,p}$ and $I$.
Inserting Eqs.\ (\ref{skin}) and (\ref{bsminus}) into
Eq.\ (\ref{stifminus}),
 one finally arrives at
\be\l{stiffin}
Q=-\nu\; \frac{J^2}{k^{}_S}, \qquad
\nu=\frac{k_S^2 I^2}{\tau^2 J^2}=
\frac{9 {\cal J}_{-}^2}{16 g^2(w_r)}\;,
\ee
where ${\cal J}_{-}$ and $g(w_r)$ are given by Eqs.\ (\ref{jminus}),
(\ref{fw}), and (\ref{boundcond}).
In the derivation of Eq.\ (\ref{stiffin}) we used
also Eq.\ (\ref{ysolplus}) for the diffuseness parameter $a$
and  Eq.\ (\ref{csymwt}) for $c_{sym}$.
Note that $Q=-9J^2/4k^{}_S$ has been predicted 
in Refs.\ \cite{myswann69,myswnp80pr96} and
therefore for $\nu=9/4$ the first part of (\ref{stiffin}), which 
relates $Q$ with
the volume symmetry energy $J$ and the isovector surface energy  $k^{}_S$
constants,
is identical to that used in Refs.\ \cite{myswann69,myswnp80pr96,myswprc77,myswiat85,vinas1,vinas2}.
However, in our derivations $\nu$ deviates from $9/4$ and it is proportional to the function
${\cal J}_{-}^2/g^2(w_r)$. This function depends significantly on the SO interaction
parameter $\beta$ but not too much on the specific Skyrme forces. Indeed, the most sensitive parameter
${\cal C}_{-}$ cancels in the expression (\ref{stiffin}) for $\nu$:
$k^{}_S \propto {\cal C}_{-} $ and
$\tau \propto 1/c_{sym}^2 \propto  {\cal C}_{-}$
[see also Eqs. \ (\ref{bsminus}) for $k^{}_S$, (\ref{csymwt})
for $c_{sym}$ and (\ref{skin}) for $\tau$].
The constant $\nu$  at $\beta= 0$ can be easy evaluated,
\be\l{nu0}
\nu \approx \frac{108}{25}\;\frac{\left[1-8/(7 c_{sym})\right]^2}{
 1-4/(3 c_{sym})},
\ee
neglecting small terms $\propto 1/c_{sym}^2$, $c_{sym}\approx 2-6$,
for the Skyrme parameters of Refs.\ \cite{chaban,reinhard}
[$c_{sym}=\infty$ for T6 forces; see Eqs.\ (\ref{stiffin}),
 (\ref{jminus}), (\ref{csymwt}), and (\ref{boundcond}) ($w_r=1/3$) and Table I].
Another difference in $Q$ [ Eq.\ (\ref{stiffin})] from that of
Refs.\ \cite{myswann69,myswnp80pr96,myswprc77,myswiat85}
 is the
expression (\ref{bsminus}) itself for
$k^{}_S$. Thus,  the isovector stiffness coefficient
$Q$ introduced originally
by Myers and Swiatecki
\cite{myswann69,myswnp80pr96} is not a parameter of our approach
but was
found analytically in the explicit closed form (\ref{stiffin})
through the parameters of the Skyrme forces.

Notice that the universal functions $w(x)$ 
[Eqs.\ (\ref{ysolplus}) and (\ref{ysolan}) ] and $w_{-}(x)$ 
[Eq.\ (\ref{ysolminus})] of the leading order in the ES approximation
can be used [explicitly analytically in the quadratic
approximation for $\epsilon(w)$] for the calculations of the surface 
energy coefficients
$b_S^{(\pm)}$ [Eq.\ (\ref{sigma})] and the neutron skin $\tau$ 
[Eq.\ (\ref{skin})]. As shown in Appendixes B and C,  
only these particle-density 
distributions  $w(x)$  and $w_{-}(x)$ within the surface layer 
are needed through their derivatives [the lower limit of the integration
over $\xi$ in Eq.\ (\ref{sigma}) can be approximately extended to
$-\infty$ because of no contributions from the internal volume region
in the evaluation of the main surface terms of the pressure and energy].
Therefore,  the surface symmetry-energy coefficient 
$k_S$ in Eqs.\ (\ref{bsminus}) and (\ref{Jm}) , the neutron skin 
$\tau$ [Eq.\ (\ref{skin})], and the 
isovector stiffness 
$Q$ [Eq.\ (\ref{stiffin})]
can be approximated analytically in terms of the functions 
of the definite critical combinations of the Skyrme parameters such as $\beta$, 
$c_{sym}$, $a$, and the parameters of the infinite nuclear matter
($b_V, \overline{\rho}, K$). Thus,  they are independent of the 
specific properties 
of the nucleus (for instance, the neutron and proton numbers), and 
the curvature and deformation of the nuclear surface in the 
considered ES approximation.  

\section{Discussion of the results}

In Table II and also in Fig.\ \ref{fig5} we show the isovector energy coefficient $k^{}_S$ [Eq.\ (\ref{bsminus})],
the stiffness parameter $Q$ [Eq.\ (\ref{stiffin})], and the neutron skin $\tau$ [Eq.\ (\ref{skin})]
obtained within the ES approximation using the quadratic approximation for $\eps(w)$  for several Skyrme forces
\cite{chaban,reinhard} with parameters presented in Table I. We also show the quantities $k_{S\;0}$, $\nu_0$, $Q_0$ and $\tau_0$
where the SO interaction is neglected ($\beta=0$). One can see a fairly good agreement for the analytical isoscalar
energy constant $b_S^{(+)}$ (\ref{bsplus}) with that of Refs.\ \cite{chaban,reinhard} and \cite{magsangzh} (Table I).
The isovector energy coefficient $k^{}_S$ is more sensitive to the choice
of the  Skyrme forces than the isoscalar one $b_S^{(+)}$
(Eq.\ (\ref{bsplus}) and Ref.\ \cite{magsangzh}).
The modulus of  $k^{}_S$  is significantly larger for most of the Lyon Skyrme forces SLy
\cite{chaban} and SkI3 \cite{reinhard} than for the other ones. For these forces
the stiffnesses $Q$ are correspondingly smaller. The isovector stiffness $Q$
is even more sensitive to the constants of the Skyrme force than the constants $k^{}_S$.
They are significantly larger for all forces, especially for SGII, than the 
well known empirical
values $Q \approx 14-35$ MeV \cite{myswprc77,myswnp80pr96,myswiat85}.

Swiatecki and his collaborators \cite{myswprc77} found the stiffness $Q \approx 14-20$ MeV by
fitting  the nuclear isovector giant dipole-resonance (IVGDR) energies calculated in
the simplest version of the hydrodynamical model to the experimental data. Later, they
suggested larger values $Q\approx 30-35$ MeV accounting for a more detailed study of
other phenomena in Refs.\ \cite{myswnp80pr96,myswiat85}.
In spite of several misprints in the derivations of the IVGDR energies in Ref.\ \cite{myswprc77}
(in particular, in Eq.\ (7.7) of \cite{myswprc77} for the displacement of the center-of-mass conservation, the
factor $NZ/A^2$ should be in the numerator and not in the
denominator of the irrotational flow moment of inertia; 
see Ref.\ \cite{eisgrei})), the final result for the IVGDR energy
constant is almost the same
as for the asymptotically large values
of $Q$, $3J A^{-1/3}/Q\ll 1$ ($NZ/A^2 \approx 1/4$),
\begin{equation}
D=\hbar \omega_{-}A^{1/3}=D_{\infty}/\sqrt{1+3J A^{-1/3}/Q},\qquad
D_\infty=\sqrt{8 \hbar^2 J/\left(mr_0^2\right)}\;.
\label{wladekGDR}
\end{equation}
These values for $D$  are in good agreement with the well known
experimental value $D_{exp}\approx 80$ MeV for heavy nuclei
($D \approx D_{\infty}\approx 88$ MeV) within a precision better than 
or of the order
of 20\% (a little worse for the specific SkI3 Skyrme forces), as shown in
Table II, see also Ref.\ \cite{denisov} for
a more proper HD approach and Refs.\ \cite{plujko1,plujko2,plujko3}
for other semiclassical nuclear models taking all into account
the nuclear surface motion. As shown in
Ref.\ \cite{BMV}, the averaged IVGDR energies and the energy weighted sum
rules (EWSR), obtained with
the semiclassical FLD approach based on the Landau-Vlasov equation
\cite{kolmagsh} with macroscopic boundary conditions (see
Appendix B), are also basically insensitive to the isovector surface energy
constant $k_S$, and they are similarly
in good agreement with the experimental data. An investigation of
the splitting of the IVGDR within this approach into the main peak
which exhausts mainly the independent-of-model EWSR  and a satellite
(with a much smaller contribution to the EWSR),  focusing on a much more
sensitive  $k_S$ dependence of the pygmy (IVGDR satellite) resonances
(see Refs. \cite{kievPygmy,nester}), will be published elsewhere.

More precise $A$ dependence of the quantity $D$ [Eq.\ (\ref{wladekGDR})]
for finite values of $Q$ seems to be beyond the accuracy of
these HD calculations because of several other reasons.
More realistic self-consistent  HF calculations accounting for the Coulomb
interaction, surface-curvature, and quantum-shell effects led to larger
$Q\approx 30-80$ MeV \cite{brguehak,vinas2}. With larger $Q$ (see Table II)
the fundamental parameter of the LDM expansion in Ref.\ \cite{myswann69},
$(9 J /4Q) A^{-1/3}$, is really small for $A \simg 40$, and therefore
 results obtained by using this expansion are more justified.

The most responsible parameter of the Skyrme HF approach leading to 
significant differences in the $k_S$ and $Q$ values
 is the constant ${\cal C}_{-}$ in the gradient terms of the energy density 
[Eq.\ (\ref{enerden}) and Table I].
Indeed, the key quantity in the expression for $Q$, Eq.\ (\ref{stiffin}), 
and the isovector surface energy constant
$k^{}_S$ [or $b_S^{(-)}$, Eq.\ (\ref{bsminus})], is the constant ${\cal C}_{-}$
because one mainly has $k^{}_S\propto {\cal C}_{-}$ (see Fig.\ \ref{fig5}), 
and $Q \propto 1/k_S \propto 1/{\cal C}_{-}$.
As seen in Table I and in Fig.\ \ref{fig5} the constant ${\cal C}_{-}$ 
is very different for the different Skyrme
forces (even in sign). As shown in Fig.\ \ref{fig3} and below 
Eq.\ (\ref{stiffin}),
other quantities in Eq.\ (\ref{stiffin}) are much less sensitive 
to most Skyrme interactions.
The situation is very much in contrast to the isoscalar energy 
density constant $ b_S^{(+)} \propto C_{+}$ [Eq.\ (\ref{bsplus})].
All Skyrme parameters are fitted to the well known experimental 
value  $b_S^{(+)} =17-19$ MeV because
${\cal C}_{+}$ is almost constant (Table I). Contrary to this,
there are so far no clear experiments which would determine $k^{}_S$
well enough because the mean energies of the IVGDR (main peaks) do not 
depend very much on $k^{}_S$
for the different Skyrme forces (see last two rows of Table II).
Perhaps, the low-lying isovector collective states are more sensitive but 
there is no careful systematic study of their
 $k^{}_S$ dependence at the present time.
Another reason for such different $k_S$ and $Q$ values might be traced back
to the difficulties in deducing $k^{}_S$ directly from the HF calculations
due to the curvature and quantum effects, in contrast to $b_S^{(+)}$.
We also have to go far away from the nuclear stability line to
subtract uniquely the coefficient $k^{}_S$ in the dependence of
$b_S^{(-)} \propto I^2=(N-Z)^2/A^2$, according to Eq.\ (\ref{bsminus}).
For exotic nuclei one has more problems to relate $k^{}_S$ to the
experimental data with a good enough precision.
Note that $k^{}_S$ is a more fundamental constant
 than the isovector stiffness $Q$ due to the direct relation to the
tension coefficient $\sigma_{-}$ of the isovector capillary pressure.
Therefore, it is simpler to analyze 
the experimental data for the IVGDR within the macroscopic HD or FLD models
in terms of the constant $k^{}_S$. The quantity $Q$ involves also the
ES approximation for
the description of the nuclear edge through the neutron skin $\tau$
[see Eq.\ (\ref{Esq})]. The precision of this
description depends more on the specific
nuclear models \cite{vinas1,vinas2,vinas3,vinas4}.
On the other hand, the neutron skin thickness $\tau$
is interesting in many aspects for the investigation of exotic nuclei,
in particular, in nuclear astrophysics.

We emphasize that for specific Skyrme forces there exists an 
abnormal behavior of the
isovector surface constants $k^{}_S$ and $Q$. It is related to the 
fundamental constant ${\cal C}_{-}$ of
the energy density (\ref{enerden}). For the parameter 
set T6  (${\cal C}_{-}=0$) one finds $k^{}_S=0$. Therefore,
according to Eq.\ (\ref{stiffin}), the value of $Q$ diverges ($\nu$ is almost independent on ${\cal C}_{-}$).
Notice that the isovector gradient terms which are important for
the consistent derivations within the ES approach are also not included
(${\cal C}_{-}=0$) in the symmetry energy density in
Refs.\ \cite{danielewicz1,danielewicz2}.
Moreover, for RATP \cite{chaban} and SkI \cite{reinhard} (also for  
the specific Skyrme forces BSk6 and BSk8 of
Ref.\ \cite{pearson}\footnote{Notice also that for the BSk forces  
\cite{pearson} we found an unexpected
behavior of the particle density $w_{-}(x)$ with a negative minimum 
near the ES (a proton instead of
 neutron skin because of $\rho_n<\rho_p$, in contrast to any other
forces discussed in Ref.\ \cite{chaban} with $w_{-}$ being always
positive).}), the isovector stiffness $Q$ is even negative 
as ${\cal C}_{-}>0$ ($k^{}_S>0$),
in contrast to all other Skyrme forces.

Table II shows also the coefficients $\nu$ of Eq.\ (\ref{stiffin}) 
for the isovector stiffness $Q$.
They are mostly constant [$\nu_0\approx 2-4$; see Eq.\ (\ref{nu0}) 
and Table II] for all Skyrme forces at $\beta=0$.
However, these constants $\nu$ are rather sensitive to the SO interaction, 
i.e., to the $\beta$
dependence of both the function $g(w_r)$ [Eq.\ (\ref{fw})] 
in expression (\ref{skin}) for the neutron skin $\tau$
and to the constant ${\cal J}_{-}$ [Eq.\ (\ref{jminus})] in 
expression (\ref{bsminus})
for the isovector energy coefficient $k^{}_S$.
As compared to 9/4 suggested in Ref.\ \cite{myswann69}, they are 
significantly smaller
in magnitude for the most of the Skyrme forces (besides those of SGII and T6 
with larger values of $\nu$).

\section{Conclusions}

Simple expressions for the isovector parts of the particle densities and 
energies in the leading ES
approximation were used for the derivation of analytical expressions of 
the surface symmetry energy, the neutron skin thickness,
and the isovector stiffness coefficients. As shown in Appendix B we have 
to include higher order terms
in the parameter $a/R$. These terms depend on the well known parameters of 
the Skyrme forces. Results for
the isovector surface energy constant $k_S$,  the neutron skin 
thickness $\tau$, and the stiffness $Q$
depend in a sensitive way on the choice of the parameters for the 
Skyrme functional, especially on the
parameter ${\cal C}_{-}$ in the gradient terms of the density in the 
surface symmetry energy density
of Eq.\ (\ref{enerden}).
The values of the isovector constants $k_S$, $\tau$, and $Q$
 depend also very much
on the SO interaction constant $\beta$. The isovector stiffness constants
$Q$ are significantly larger than those found earlier for all desired
Skyrme forces.
The mean IVGDR energies and sum rules calculated
in the HD \cite{myswprc77,denisov,eisgrei}  and FLD \cite{kolmagsh,BMV} models
for most  of the $k_S $ values in Table II
are in a fairly good agreement with the
experimental data. For further perspectives, it would be worthwhile to
apply our results to the calculations of the pygmy resonances in the IVGDR
strength
within the FLD model \cite{kolmagsh}
and the isovector low-lying collective states within the periodic orbit
theory \cite{gzhmagfed,blmagyas,BM}, which are expected to be more sensitive
to the values of $k_S$. Our approach is helpful for further study of the
effects in the surface symmetry energy because it gives the analytical
universal expressions for the constants $k^{}_S$, $\tau$ and $Q$ 
which are independent of the specific properties of the nucleus. 
These constants are directly connected with
a few critical parameters of the Skyrme interaction without using
any fitting.

\bs
\centerline{{\bf Acknowledgements}}
\ms

The authors thank
M. Brack, M. Brenna, V.Yu. Denisov, V.M. Kolomietz, J. Meyer, V.O. Nesterenko,
M. Pearson, V.A. Plujko, X. Roca-Maza, A.I. Sanzhur,
and X. Vinas for many useful discussions. One of us (A.G.M.) is
also very grateful for hospitality during his working visits to the
National Centre for Nuclear Research in Otwock-Swierk of Poland.
This work was partially supported by the Deutsche
Forschungsgemeinschaft Cluster of Excellence 'Origin and
Structure of the Universe' (www.universe-cluster.de).

\appendix \setcounter{equation}{0}
\renewcommand{\theequation}{A\arabic{equation}}

\begin{center}
\textbf{Appendix A: Solutions to the isovector Lagrange equation}
\label{appA}
\end{center}

The Lagrange equation for the variations of the isovector particle 
density $\rho_{-}$
in the energy density (\ref{enerden}) up to the leading terms in a 
small parameter
$a/R$ is given by \cite{magsangzh}
\be\l{lagrangeqminus0}
{\cal C}_{-} \; \frac{\partial^2 \rho_{-}}{\partial \xi^2} +
\frac{\d}{\d \rho_{-}} \left[\rho_{+} \epsd_{-}(\rho_{+},\rho_{-})\right]= 0,
\ee
where $\epsd_{-}$ is defined just below Eq.\ (\ref{enerden}). We neglected here
the higher order terms proportional to the first derivatives of the
particle
density $\rho_{-}$ with respect to $\xi$ and
the surface correction to the isovector chemical potential
as in Refs.\ \cite{strmagbr,strmagden} for the isoscalar case.
For the dimensionless isovector density
$w_{-}=\rho_{-}/(\overline{\rho} I)$,
after simple transformations one finds the equation and the
 boundary condition in the form
\be\l{yeq0minus}
\frac{\d w_{-}}{\d w} =c_{sym}
\sqrt{\frac{1+ \beta w}{\eps(w)}}
\;\sqrt{1 -\frac{w_{-}^2}{w^2}}, \qquad w_{-}(1) =1\;,
\ee
where $\beta$ is the SO parameter defined below Eq.\ (\ref{ysolplus});
see also Eq.\ (\ref{csymwt}) for $c_{sym}$.
The above equation determines the isovector density $w_{-}$ as a function of
the isoscalar one $w(x)$ [Eq.\ (\ref{ysolplus})]. In the
quadratic approximation for $\eps(w)$ one explicitly finds
\bea\l{ysolan}
x(w)&=&\sqrt{1+\beta}\,\ln\left[
\frac{(1-w) \left(1+(1+2\beta)w_r + 2 \sqrt{(1+\beta)(1+\beta w_r)w_r }
\right)}{(1-w_r) \left(1+(1+2\beta)w + 2 \sqrt{(1+\beta)(1+\beta w)w }
\right)}\right]
\nonumber\\
&+& \sqrt{-\beta}\left[\arcsin\left(1+2 \beta w\right)-
\arcsin\left(1+2 \beta w_r\right)\right]\;, \qquad x < x(w=0)
\eea
and $w=0$ for $x \geq x(w=0)$;  $w_r$ is the solution of 
the boundary condition (\ref{boundcond}).
Substituting
$w_{-}=w\;\cos \psi$ into Eq.\ (\ref{yeq0minus}),
and taking the approximation $\eps=(1-w)^2$,
one has the following
first order differential equation for a new function $\psi(w)$:
\be\l{ueqminus}
\frac{w(1-w)}{c_{sym}}\;\sin \psi\;\frac{\d \psi}{\d w} = 
\sqrt{1+\beta w}\; \sin\psi -
\frac{1-w}{c_{sym}}\;\cos\psi,
\qquad\qquad \psi(1)=0.
\ee
The boundary condition for this equation is related to that
of Eq.\ (\ref{yeq0minus}) for $w_{-}(w)$.
This equation looks more complicated because of the
trigonometric nonlinear terms.
However, it allows us to obtain simple approximate and rather exact
analytical solutions within standard perturbation theory.
Indeed, according to Eqs.\ (\ref{yeq0minus}) and (\ref{ysolplus})
where we did not express the $x$ dependence explicitly,
we note that $w_{-} \propto w(x)$ is a sharply decreasing function of
$x$ within a small diffuseness region of the
order of 1 in dimensionless units (Figs.\ \ref{fig2}-\ref{fig4}). Thus,
 we may find the approximate solutions to the equation (\ref{ueqminus})
(with its boundary condition), in terms
of a power expansion of a new function
$\widetilde{\psi}(\widetilde{w})$ in terms of a  new small
argument $\widetilde{w}$,
\be\l{powser}
\widetilde{\psi}(\widetilde{w})\equiv
\psi(w)=\sum_{n=0}^{\infty} c_n\;\widetilde{w}^n\;,
\ee
with unknown coefficients $c_n$ and $\widetilde{w}$ defined in
Eq.\ (\ref{csymwt}). Substituting the power series
(\ref{powser}) into Eq.\ (\ref{ueqminus}) one expands first
the trigonometric functions into power series of $\widetilde{w}$ in
accordance with the boundary condition in Eq.\ (\ref{ueqminus}).
As usual, using standard perturbation theory,
we obtain the system of algebraic equations
for the coefficients $c_n$ [Eq.\ (\ref{powser})] by
equating the coefficients from both sides of
Eq.\ (\ref{ueqminus}) at the same powers of $\widetilde{w}$. This
simple procedure leads to a system of algebraic recurrence relations
which determine the coefficients $c_n$ as functions of the parameters
$\beta$ and $c_{sym}$ of
Eq.\ (\ref{ueqminus}),
\bea\l{cn}
c_0&=&0,\qquad\qquad c_1=\frac{1}{\sqrt{1+\beta}},\qquad\qquad
c_2=\frac{c_1}{\sqrt{1+\beta}}\left(
\frac{\beta c_{sym} }{2 \sqrt{1+\beta}}-c_1\right),
\nonumber\\
c_3&=&\frac{1}{\sqrt{1+\beta}}\left\{
\frac{\beta^2 c_{sym}^2 c_1}{8 \left(1+\beta\right)^{3/2}} +
c_1^2\left(c_{sym}-\frac12\right) + \frac16 \sqrt{1+\beta} c_1^3
+
c_2\left(\frac{\beta c_{sym}}{2 \sqrt{1+\beta}}
-3 c_1\right)\right\},
\eea
etc. In particular, up to the second order in $\widetilde{w}$,
we derive an analytical solution in an explicitly closed form:
\be\l{wsol3}
\widetilde{\psi}(\widetilde{w}) = \widetilde{w}\left(c_1 +
c_2 \widetilde{w}\right),\quad
c_1=\frac{1}{\sqrt{1+\beta}},\quad
c_2=\frac{1}{(1+\beta)^{3/2}}
\left(\frac{\beta}{2} c_{sym}-1\right).
\ee
Thus,  using the standard
perturbation expansion method of solving
$\widetilde{\psi}(\widetilde{w})$ in terms of the power series
of the $\widetilde{w}$ (up to $\widetilde{w}^2$), one obtains
the quadratic expansion of $\psi(w)$ [Eq.\ (\ref{ysolminus})]
with $\widetilde{c}=c_2/c_1$.
Notice that one finds a good convergence of the power expansion
of $\widetilde{\psi}(\widetilde{w})$ (\ref{wsol3}) in $\widetilde{w}$ for
$w_{-}(x)$ at the second order in $\widetilde{w}$ because of the large
value of $c_{sym}$ for all Skyrme forces presented in Table I [Eq.\ (\ref{csymwt}) for $c_{sym}$].

\appendix \setcounter{equation}{0}
\renewcommand{\theequation}{B\arabic{equation}}

\begin{center}
\textbf{Appendix B: The macroscopic boundary conditions and surface
tension coefficients}
\label{appB}
\end{center}

For the derivation of the expression for the surface tension coefficients
$\sigma_{\pm}$, we first write the system of the Lagrange equations
by using variations of the energy density $\epsi(\rho_{+},\rho_{-})$
with respect to the isoscalar and isovector densities $\rho_{+}$
and $\rho_{-}$. Then, we substitute the solution of the first Lagrange equation
for the variations of the
isoscalar density $\rho=\rho_{+}$ in the energy density
(\ref{enerden}) (Refs.\ \cite{strmagbr,strmagden}) into the second
Lagrange equation for the isovector density $\rho_{-}$. Using
the Laplacian in  the variables $\xi$ and $\eta$ (Appendix A in Ref.\
\cite{strmagbr}) we keep the major terms in this second equation
within the improved precision in the small parameter
$a/R$. The improved precision means that we take into account
the next terms proportional to
the first  derivatives of the particle densities [along with
the second ones of Eq.\ (\ref{lagrangeqminus0})] and the
small surface
corrections $\Lambda_{\pm}$ to the
isoscalar and isovector Lagrange multipliers $\lambda_{\pm}$.
Within this improved precision,
one finds the second Lagrange equation by the variations of the energy
density $\epsi(\rho_{+},\rho_{-})$, Eq.\ (\ref{enerden}), with respect to
 the isovector
particle density $\rho_{-}$:
\be\l{lagrangeqminus1}
{\cal C}_{-} \; \frac{\partial^2 \rho_{-}}{\partial \xi^2} +
2 {\cal C}_{-} H \frac{\partial \rho_{-}}{\partial \xi} -
\frac{\d}{\d \rho_{-}} \left[\rho_{+} \epsd_{-}(\rho_{+},\rho_{-})\right] +
\Lambda_{-} = 0,
\ee
where $H$ is the mean curvature of the ES ($H=1/R$ for the spherical ES). The
isovector chemical-potential correction $\Lambda_{-}$ was
introduced \cite{magsangzh}
like the isoscalar one $\Lambda_{+}$, worked out in detail in
Refs.\ \cite{strmagbr,strmagden}.
Multiplying Eq.\ (\ref{lagrangeqminus1})
by $\partial \rho_{-}/\partial \xi$ we integrate in the  coordinate $\xi$
normal to the ES from a spatial point $\xi_{in}$
inside the volume (at $\xi_{in} \siml -a$ )
to $\infty$ term by term.
Using also integration by parts, within the ES approximation
this results in the macroscopic boundary conditions (together with the isoscalar
condition from
\cite{strmagbr,strmagden,magsangzh,kolmagsh,magstr,magboundcond})
\be\l{macboundcond}
\left(\overline{\rho}\;I\;
\Lambda_{-}\right)_{ES} =  P_{s}^{(-)} \equiv 2 \sigma_{-} H, \qquad
 \left(\overline{\rho}\;
\Lambda_{+}\right)_{ES} =  P_{s}^{(+)} \equiv 2 \sigma_{+} H.
\ee
Here, $P_s^{(\pm)}$ are the
isovector and isoscalar surface-tension (capillary) pressures and
 $\sigma_{\pm}$ are the corresponding tension coefficients;
see their expressions in Eq.\ (\ref{sigma}). We point out that the lower limit
$\xi_{in}$ can be approximately extended to $-\infty$ as in Eq.\ (\ref{sigma})
for $\sigma_\pm$. The integrands contain the square of the first derivatives, 
$(\partial \rho_{\pm}/\partial \xi)^2 \propto (R/a)^2$, 
and the integral over $\xi$ 
converges exponentially rapidly within the ES layer $|\xi| \leq a$. This leads
to the aditional small factor $a/R$ in Eq.\ (\ref{sigma}), 
$\sigma_{\pm} \propto R/a$. Therefore, at this higher order of the improved
ES approximation one may neglect higher order corrections 
in the calculation
of derivatives of $\rho_{\pm}$ themselves by using the analytical universal
density distributions $w_{\pm}(x)$ [Eqs.\ (\ref{ysolplus}), (\ref{ysolan}) 
and (\ref{ysolminus})] within the ES layer which do not depend on the specific
properties of the nucleus as mentioned in the main text. 
(These corrections are small terms proportional to the first derivative
$\partial \rho_{-}/\partial \xi$ and $\Lambda_{-}$ in
Eq.\ (\ref{lagrangeqminus1}),
as for the isoscalar
case considered in
Refs.\ \cite{strmagbr,strmagden,magstr,magboundcond}).
In these derivations,  the obvious boundary
conditions of disappearance of the particle densities and all their
derivatives with respect to $\xi$ outside of the ES for $\xi \rightarrow \infty$
($\xi \gg a$) were taken into account too.

The Lagrange multipliers $\Lambda_{\pm}$ multiplied by
$\overline{\rho}I$ and $\overline{\rho}$ in the parentheses
on the left-hand sides
 of equations (\ref{macboundcond})
are the volume isovector ($\overline{\rho}I \Lambda_{-}$) and isoscalar
($\overline{\rho} \Lambda_{+}$) pressure excesses, respectively
(see Ref.\ \cite{magsangzh}).
These pressures due to the surface curvature can be derived using the
volume solutions of
the Lagrange equations for the particle densities [obtained by doing
variations of the energy density $\epsi$ and neglecting
all the derivatives of the particle densities in Eq.\ (\ref{enerden})],
\be\l{rhovol2}
\rho_{-} \approx \overline{\rho}\; \left[I\left(1 +
\frac{9 \Lambda_{+}}{K}\right) +
\frac{\Lambda_{-}}{2J} \right],\quad \rho_{+}\approx
\overline{\rho}\left[1+\frac{9 \Lambda_{+}}{K}
\left(1 - \frac{81\Lambda_{+)}}{2K}\right) -
\frac{18 J}{2K}\;I^2\right]\;.
\ee
Inserting $\Lambda_{+}$ and $\Lambda_{-}$ from Eq.\ (\ref{macboundcond})
into Eq.\ (\ref{rhovol2}) one gets
\be\l{rhovolbsminus2}
\rho_{-}=\overline{\rho}\;I\;
\left[1 + \frac{6 b_{S}^{(+)}\;H\;r_0}{K} +
\frac{2 b_{S}^{(-)}\;H\;r_0}{6 J\;I^2}\right].
\ee
As seen from Eq.\ (\ref{rhovolbsminus2}),
 the isovector density correction to the volume density
$\rho_{-}$ due to a finiteness
of the coupled system of the two Lagrange equations depends on both
isoscalar and isovector
surface energy constants $b_S^{(\pm)}$ in the first-order expansion of the small
parameter $a/R$. If we are not too far from the valley of stability $I$ is an
additional small parameter and the isovector corrections are small compared
with the isoscalar values
[$b_S^{(-)} \propto I^2$, $\Lambda_{-}\propto I$;
see Eqs.\ (\ref{rhovol2}), (\ref{rhovolbsminus2}), and (\ref{bsminus})].
Thus,  Eq.\ (\ref{macboundcond}) has  a clear
physical meaning as the macroscopic boundary conditions
for equilibrium of the isovector and isoscalar
forces (volume and surface pressures) acting on the ES \cite{bormot,kolmagsh}.
Note that the isovector tension coefficient  $\sigma_{-}$
is much smaller than the isoscalar one $\sigma_{+}$
[see Eq.\ (\ref{sigma})] as
$\sigma_{-} \propto  I^2$ due to $\rho_{-} \propto I$ and $I \ll 1$
is small near the nuclear stability line. Another reason is
the smallness of ${\cal C}_{-}$
as compared to ${\cal C}_{+}$ for the realistic Skyrme forces
\cite{chaban,reinhard}.
From comparison of Eqs.\ (\ref{rhovol2})  and
(\ref{rhovolbsminus2}) for $\rho_{-}$ [see also Eq.\ (\ref{sigma})],
one may also evaluate
\be\l{lambdatotminus}
\Lambda_{-} =
\frac{2 \sigma_{-} H}{\overline{\rho}\;I} \approx
\frac{2 b_S^{(-)}}{3 I A^{1/3}}
\sim k_S I \frac{a}{R}\;.
\ee
which is consistent with Eq.\ (\ref{lagrangeqminus1})
($r_0 H \sim a/R$ in these estimations, see
corresponding ones in Refs.\ \cite{strmagbr,strmagden}).

\appendix \setcounter{equation}{0}
\renewcommand{\theequation}{C\arabic{equation}}

\begin{center}
\textbf{Appendix C: Derivations of the surface energy and
its coefficients}
\label{appC}
\end{center}

\begin{figure}
\begin{center}
\includegraphics[width=0.6\textwidth,clip]{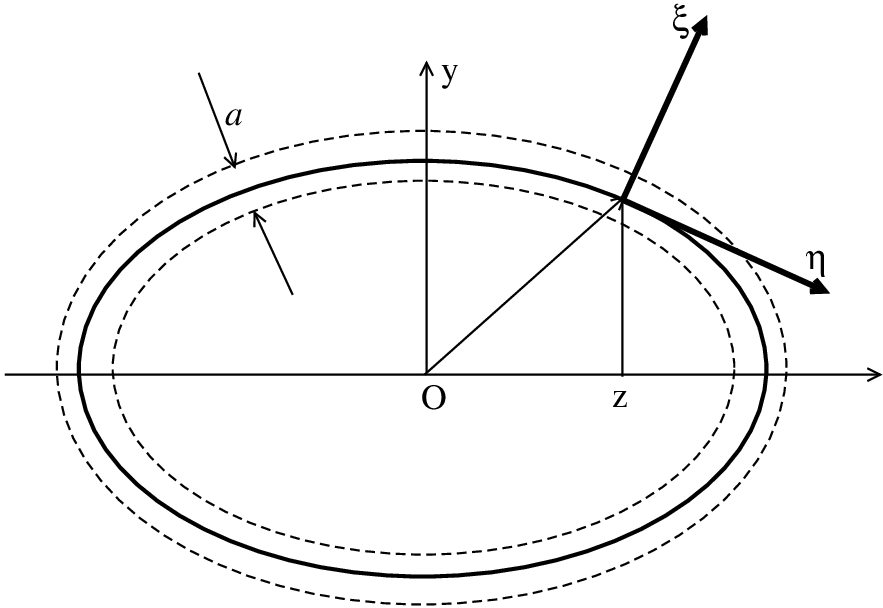}
\end{center}

\vspace{-0.2cm}
\caption{The ES and local  $\xi,\eta$ coordinates.
The ES 
in the cylindrical $y,z$ coordinates with the symmetry axis $z$ and
diffuseness parameter $a$ are shown schematically by
the thick solid and dashed curves
(after Ref.\ \cite{magsangzh}).} 
\label{fig1}
\end{figure}

\begin{figure}
\begin{center}
\includegraphics*[width=0.70\textwidth]{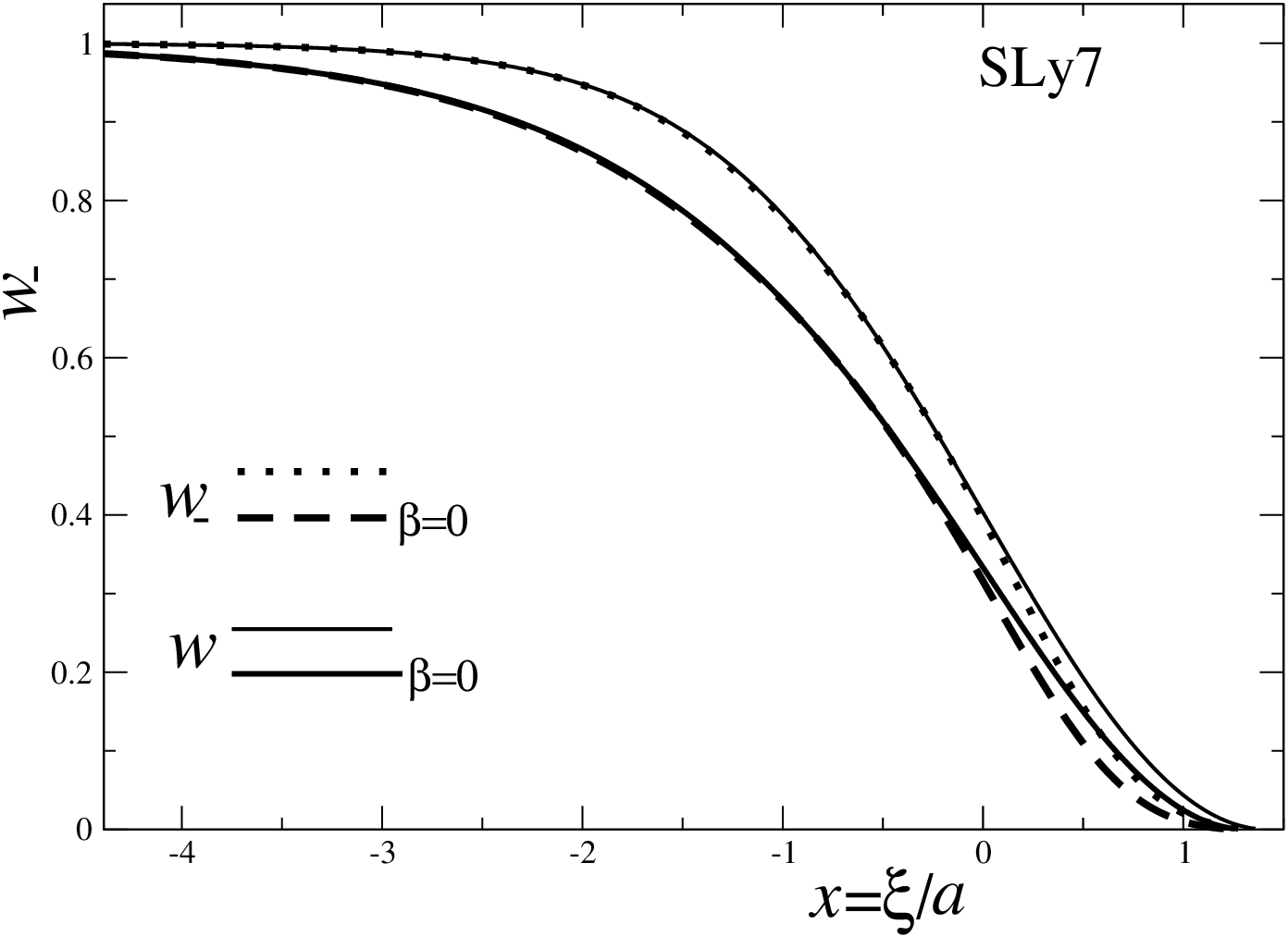}
\end{center}

\vspace{-0.2cm}
\caption{ Isovector $w_{-}$
[Eq.\ (\ref{ysolminus})] particle density vs $x=\xi/a$ with
and without ($\beta=0$) SO terms for the Skyrme force SLy7 as a typical
example
like in Ref.\ \cite{magsangzh}; the isoscalar $w$ 
[see Eqs.\ (\ref{ysolplus}), (\ref{ysolan})] is also shown by solid lines.}

\label{fig2}
\end{figure}

\begin{figure}
\begin{center}
\includegraphics[width=0.70\textwidth]{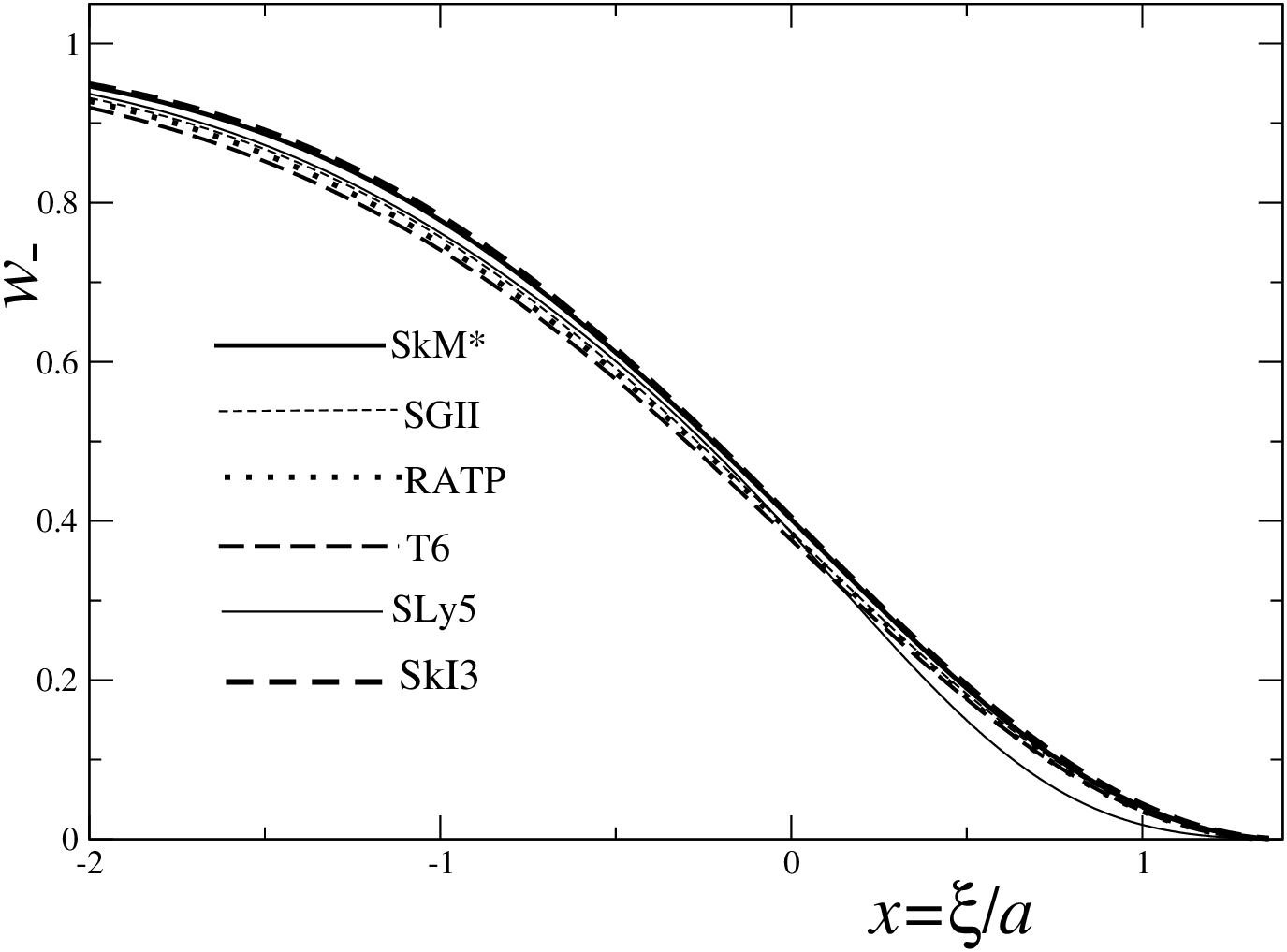}
\end{center}

\vspace{-0.2cm}
\caption{Isovector particle densities
$w_{-}(x)$ (\ref{ysolminus})  as functions of $x$ within the
quadratic approximation to
$\eps(w)$
for several Skyrme forces \cite{chaban,reinhard}, 
see also Ref.\ \cite{magsangzh}.}

\label{fig3}
\end{figure}

\begin{figure}
\begin{center}
\includegraphics[width=0.70\textwidth]{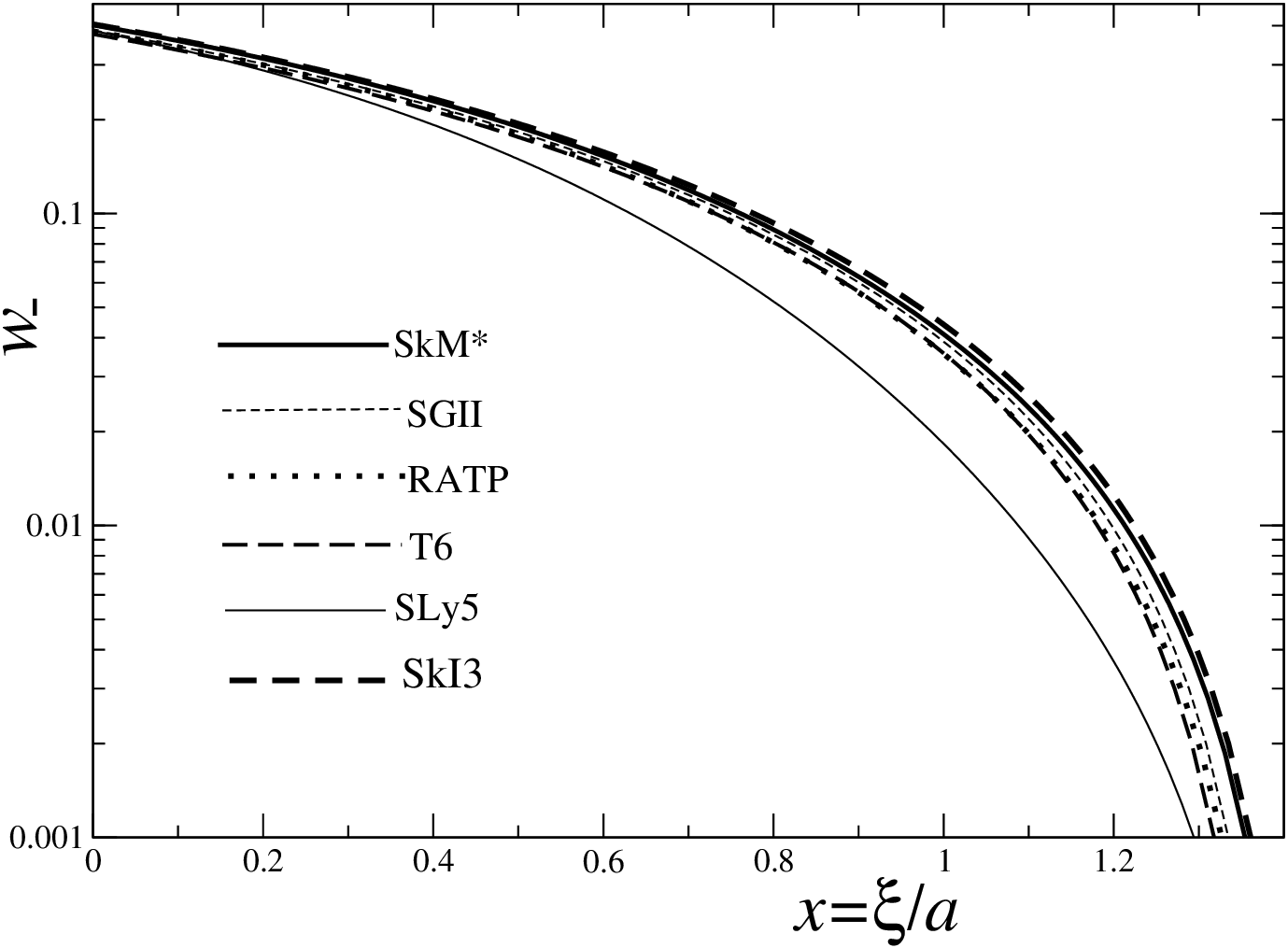}
\end{center}

\vspace{-0.2cm}
\caption{The same as in Fig.\ \ref{fig3} for the critical
Skyrme interactions but within a smaller
edge diffuseness  region in the logarithmic scale.}

\label{fig4}
\end{figure}

\begin{figure}
\begin{center}
\includegraphics*[width=0.70\textwidth]{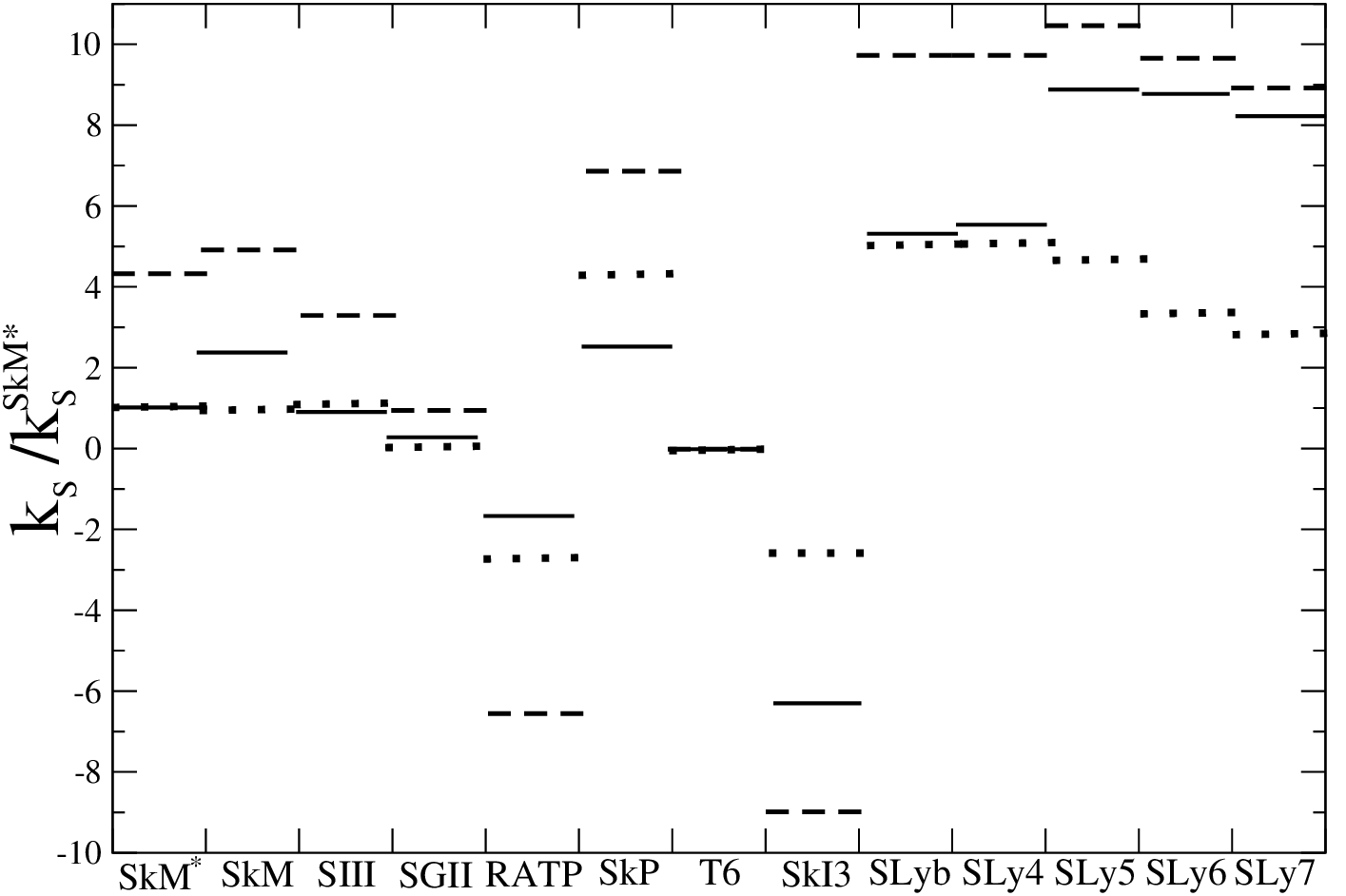}
\end{center}

\vspace{-0.2cm}
\caption{Isovector energy constant $k^{}_S$ of Eq.\ (\ref{bsminus})
(solid bars) vs the coefficient
${\cal C}_{-}$ (dotted points) of
Eq.\ (\ref{enerden}) normalized both by their values $k_S^{SkM^*}$ 
and ${\cal C}_{-}^{SkM^*}$ for the Skyrme
force SkM$^*$, respectively; $k_{s,0}$ (dashed bars) in the same units 
$k_S^{SkM^*}$ 
is given by Eq.\
(\ref{bsminus}) without the spin-orbit interaction ($\beta=0$).}

\label{fig5}
\end{figure}

For calculations of the surface energy components
$E_S^{(\pm)}$ of the energy $E$, Eq.\ (\ref{energy}),
within the same improved ES approximation as described above
in Appendix B, we first may
separate the volume terms related
to  the first two terms of Eq.\ (\ref{enerden}) for the energy density
$\epsi$. Other terms of the energy density
 $\epsi(\rho_{+},\rho_{-})$ in
Eq.\ (\ref{enerden}) lead to the surface components $E_S^{\pm}$
[Eq.\ (\ref{Espm})], as they
are concentrated near the ES.
Integrating the energy density $\epsi$ [see Eq.\ (\ref{enerden})]
over the spatial coordinates
$\r$ in the local coordinate system $\xi,\eta$ (see Fig.\ \ref{fig1}) in the ES
approximation, one finds
\be\l{Espm1}
E_S^{(\pm)} =
{\cal C}_{\pm} \oint  \d {\cal S} \int_{\xi_{in}}^{\infty} \d \xi
\left[\left(\nabla \rho_{\pm}\right)^2 -
\rho_{+}\epsd_{\pm}\left(\rho_{+},\rho_{-}\right)\right]
\approx \sigma_{\pm}\;{\cal S}\;,
\ee
where $\xi_{in} \siml -a$ is as in Appendix B
\cite{strmagbr,strmagden,magsangzh}.
The local coordinates $\xi,\eta$
were used because the integral
over $\xi$ converges rapidly within the ES layer which is effectively taken
for $|\xi|\siml a$. Therefore, again, we may extend
formally $\xi_{in}$ to $-\infty$ in the first (internal) integral
taken over the ES in the normal direction $\xi$ in
Eq.\ (\ref{Espm1}).
Then, the second integration is performed over the closed surface
of the ES. The integrand over $\xi$ contains terms of the order of
$(\overline{\rho}/a)^2 \propto (R/a)^2$ as the ones of the leading order
in Eq.\ (\ref{lagrangeqminus1}) [see for instance the second derivatives
in Eq.\ (\ref{lagrangeqminus0}) which are also $\propto (R/a)^2$].
However, the integration is effectively performed over the edge region
of the order of $a$ that leads to
the additional smallness proportional to $a/R$ as in Appendix B.
At this leading order the $\eta$ dependence of the internal
integrand can be neglected.
Moreover, from the Lagrange equations at this order one can realize that
the terms without the
particle density gradients in Eq.\ (\ref{Espm1}) are equivalent to the gradient
terms. Therefore, for the calculation
of the internal integral we may approximately reduce the integrand
over $\xi$ to the only derivatives of the universal particle densities
of the leading order $\rho_{\pm}(\xi)$ in $\xi$
(with the factor 2) using
$\left(\nabla \rho_{\pm}\right)^2-
\rho_{+}\epsd_{\pm}\left(\rho_{+},\rho_{-}\right) \approx
2(\partial \rho_{\pm}/\partial \xi)^2$
[see Eqs.\ (\ref{ysolplus}) and (\ref{ysolminus}) for $w_{\pm}(x)$] . Taking
the integral over $\xi$ within the infinite
integration region ($-\infty <\xi <\infty$) off the integral over
the ES ($\d {\cal S}$) we are left with the
integral over the ES itself that is the
surface area ${\cal S}$. Thus,
we arrive finally at the right-hand side of Eq.\ (\ref{Espm1})
with the surface tension coefficient $\sigma_{\pm}$
[Eq.\ (\ref{sigma})].

Using now the quadratic approximation $\eps(w)=(1-w)^2$
in Eq.\ (\ref{sigma}) for $b_S^{(\pm)}=4 \pi r_0^2 \sigma_{\pm}$
($|{\cal D}_{-}/ {\cal D}_{+}| \ll 1$),
one obtains (for $\beta<0$, see Table I)
\be\l{bsJpm}
b_S^{(\pm)}= 6 \overline{\rho}\; {\cal C}_{\pm}\;{\cal J}_{\pm}/(r_0 a)\;,
\ee
where
\bea\l{Jp}
{\cal J}_{+}&=&\int_0^1 \d w\; \sqrt{w(1+\beta w)}\;(1-w)
=\frac{1}{24}\;(-\beta)^{-5/2}\;
\left[{\cal J}_{+}^{(1)}\; \sqrt{-\beta(1+\beta)}\right.
\nonumber\\
&+&\left. {\cal J}_{+}^{(2)}\; \arcsin\sqrt{-\beta}\right],\qquad
{\cal J}_{+}^{(1)}=3 + 4 \beta(1+\beta),\qquad
{\cal J}_{+}^{(2)}=-3-6\beta\;.
\eea
For the isovector energy constant ${\cal J}_{-}$ one finds
\bea\l{Jm}
{\cal J}_{-}&=&-\frac{1}{1+\beta}\;
\int_0^1 \d w\; \sqrt{
w(1+\beta w)}\;(1-w)(1+\widetilde{c} \widetilde{w})^2
\nonumber\\
&=&\frac{\widetilde{c}^2}{1920 (1+\beta) (-\beta)^{9/2}}\;
\left[{\cal J}_{-}^{(1)}\left(c_{sym}/\widetilde{c}\right)\;
\sqrt{-\beta(1+\beta)} +
{\cal J}_{-}^{(2)}\left(c_{sym}/\widetilde{c}\right)\;
\arcsin\sqrt{-\beta}\right],
\nonumber\\
{\cal J}_{-}^{(1)}(\zeta)&=& 105- 4 \beta \left\{95 +75 \zeta +
\beta \left[119+10\zeta (19+6\zeta) +8 \beta^2
\left(1+ 10\zeta(1+\zeta)\right)\right.\right.
\nonumber\\
&+&
\left.\left. 8 \zeta \left(5 \zeta (3 +2 \zeta) -6\right)\right]\right\},
\nonumber\\
{\cal J}_{-}^{(2)}(\zeta)&=&15 \left\{7+2\beta \left[
5 (3 + 2 \zeta) + 8 \beta (1+\zeta)
\left(3 +\zeta +2 \beta (1+\zeta)\right)\right]\right\}.
\eea
These equations determine explicitly the analytical expressions for the
isoscalar ($b_S^{(+)}$) and isovector ($b_S^{(-)}$) energy constants in terms of
the Skyrme force parameters; see
Eqs.\ (\ref{ysolminus}) for $\widetilde{c}$, (\ref{csymwt})
for $c_{sym}$ and $\widetilde{w}$.
For the limit $\beta \rightarrow 0$ from Eqs.\ (\ref{Jp}) and (\ref{Jm})
one has
${\cal J}_{\pm} \rightarrow 4/15$. With Eqs.\
(\ref{skin}) and (\ref{fw}) one arrives also
at the explicit analytical expression  for the isovector stiffness $Q$
as a function of ${\cal C}_{-}$ and $\beta$. In the limit
${\cal C}_{-} \rightarrow 0$  one obtains $k^{}_S \rightarrow 0$ and
$Q \rightarrow \infty$ because of the finite limit of
the argument $c_{sym}/\widetilde{c}\rightarrow (1+\beta)/\beta$
of the function ${\cal J}_{-}$ in Eq.\ (\ref{Jm})
[see also Eqs.\ (\ref{ysolminus}) for
$\widetilde{c}$ and (\ref{csymwt})
for $c_{sym}$].

\newpage
\vspace{-2.0cm}
%

%
\noindent
\hspace{1cm}
\begin{table*}[pt]
\begin{tabular}{|c|c|c|c|c|c|c|c|c|c|c|c|c|c|}
\hline
 & SkM$^*$ & SkM & SIII & SGII & RATP &SkP &
T6 & SkI3 & SLyb & SLy4 & SLy5 & SLy6 & SLy7\\
\hline
$\overline{\rho} $ (fm$^{-3}$)  & 0.16 & 0.16 &  0.15 &  0.16 & 0.16 & 0.16
  & 0.16 & 0.16 & 0.16 & 0.16 & 0.16 & 0.17 & 0.16  \\
$b^{}_{V}$ (MeV) & 15.8& 15.8 &  15.9&   15.6 & 16.0 & 15.9
   & 16.0 & 16.0 & 16.0 & 16.0 & 16.0 & 16.0 & 16.0  \\
$K$ (MeV) & 217& 217 &  355&   215& 240 & 201
   & 236 & 258 & 230 & 230 & 230 & 230 & 230   \\
$J$ (MeV)  & 30.1 & 31.0 &  28.2 & 26.9 & 29.3& 30.0
& 30.0 & 34.8 & 32.0 & 32.0 & 32.1& 31.9 & 31.9  \\
${\cal C}_{+}$ (MeV$\cdot$fm$^{5}$) & 57.6 & 52.9 &   49.4&   43.9 & 60.2 &60.1
 & 55.1 & 51.8& 59.5 & 59.5 & 59.3& 54.0 & 52.7  \\
${\cal C}_{-}$ (MeV$\cdot$fm$^{5}$) & -4.79 &-4.69 & -5.59 & -0.94 & 13.9  &-20.2
 &  0 & 12.6 & -22.3 & -22.3 & -22.8 & -15.6 & -13.4 \\
$a$ (fm) & 0.52 &0.50 & 0.59 & 0.45& 0.55 & 0.50
   &0.52 & 0.53 & 0.53 & 0.53 & 0.53 & 0.52 & 0.50 \\
$a^{}_d$ (fm) & 0.63 & 0.58 & 0.73 & 0.58 & 0.71 & 0.71
  & 0.70 &  0.63 & 0.68 & 0.68 & 0.68 & 0.61 & 0.61 \\
$c_{sym}$ & 3.26 & 3.21 & 3.42 & 6.02& 2.00 & 1.52
   &$\infty $ & 2.20 & 1.59 & 1.59 & 1.57 & 1.80 & 1.93 \\
$\beta$ & -0.64 &-0.69 &  -0.57 & -0.54 & -0.52 & -0.37
   &-0.45 & -0.65 & -0.55 & -0.55 & -0.58 & -0.59 & -0.65 \\
$b_{S}^{+}$ (MeV) \cite{chaban,reinhard}& 16.0& 16.0 & 17.0 & 14.8 & 17.9 & 17.9
& 17.9 & 16.0 & 16.7 & 18.1& 18.0 & 17.4 & 17.0 \\
$b_{S}^{+}$ (MeV) & 21.2 &19.9&  14.5 & 18.7 & 21.7 & 24.9
& 21.3 & 18.3 & 21.7 & 21.7 & 21.5 & 21.6 & 19.6  \\
\hline
\end{tabular}

\vspace*{1.0cm}
TABLE\ I. Basic parameters of the Skyrme forces from
Refs.\ \cite{chaban,reinhard}
and the isoscalar surface energy constants  $b_S^{(+)}$ of Eq.\ (\ref{bsplus});
the critical parameters $a$ [Eq.\ (\ref{ysolplus})] and $a_d$ [Eq.\ \ref{ad}]
for the nuclear diffuseness edge,
the isoscalar and isovector constants $C_{\pm}$ of the energy density
[Eq.\ (\ref{enerden})],
$c_{sym}$  [Eq.\ (\ref{csymwt})] and the spin-orbit constant $\beta$
[see below Eq.\ (\ref{ysolplus})];
SLyb denotes shortly SLy230b of Ref.\ \cite{chaban}.
\end{table*}

\newpage
%
\vspace{-2.0cm}

\newpage

\newpage
\vspace{1.5cm}

\newpage
%

\newpage
%
\noindent
\hspace{1cm}
\begin{table*}[pt]
\begin{tabular}{|c|c|c|c|c|c|c|c|c|c|c|c|c|c|}
\hline
 & SkM$^*$ & SkM & SIII & SGII & RATP &SkP &
T6 & SkI3 & SLyb & SLy4 & SLy5 & SLy6 & SLy7\\
\hline
$k_{S,0}$ (MeV)& -3.26 & -3.84 & -2.65 & -0.71 & 5.25 & -5.36
         & 0 & 6.93 & -7.51  & -7.54 & -8.14  & -7.45 &-6.95  \\
$k^{}_S$ (MeV) & -0.77 &-1.90 & -0.52 & -0.21 & 1.42 & -1.93
    & 0  & 4.88 & -4.24  & -4.38 & -6.96 & -6.72 & -6.32 \\
$\nu_0$  & 3.08 & 3.06 & 3.14 & 3.64 & 2.38 & 2.17
         & 4.32  & 2.53 & 2.12  & 2.12 & 2.12  & 2.22 & 3.32  \\
$\nu$  & 0.34 & 0.46 & 1.42 & 17.9 & 0.45 & 1.76
         & 4.30  & 0.56 & 0.44  & 0.44 & 0.59  & 0.65 & 0.67  \\
$Q_0$ (MeV)  & 7744 & 9487 & 6255 & 16879 & -371 & 3815
   & $\infty$ & -6314 & 4771  & 4794 &  5178  & 5350 & 5703   \\
$Q$ (MeV)  &   398 & 234 & 2168 & 60998 & -270 & 823
   & $\infty$ & -140 & 105  & 104 & 87  & 98 & 109   \\
$\tau_0/I$ & 0.021 & 0.020 & 0.021 & 0.0065 & 0.038 & 0.037
   & 0 & 0.033 & 0.040 & 0.40 & 0.040 & 0.037 & 0.035  \\
$\tau/I$ & 0.044 & 0.090 & 0.015 & 0.0019 & 0.072 & 0.048
   & 0 & 0.187 & 0.20 & 0.21 & 0.28 & 0.26 &  0.24  \\
$D_{HD}$ (MeV) & 85 86 & 85 86 & 82 & 82 & 90 89 & 87
   & 88 & 105 100 & 81 84 & 81 84 & 79 83 & 81 85 &  81 84  \\
$D_{FLD}$ (MeV) & 73 82 & 71 76 & 79 104 & 74 77 & 77 87 & 70 69
   & 86 88 & 101 106 & 80 90 & 80 90  & 76 84 & 80 91 &  77 89  \\
\hline
\end{tabular}

\vspace{0.5cm}
TABLE\ II. Isovector energy  $k^{}_S$
and stiffness $Q$ coefficients are shown for several Skyrme forces
\cite{chaban,reinhard}; $\nu $
is the constant of Eq.\ (\ref{stiffin});
$\tau/I$ is the neutron skin
calculated by Eq.\ (\ref{skin});
quantities $k_{S,0}$, $\nu_0$, $Q_0$ and $\tau_0$ are calculated
with $\beta=0$;
the intervals of monotonic functions
$D_{HD}(A)$ and $D_{FLD}(A)$ for the HD and FLD models in the last two lines
are related to $A\approx 50-200$ (the last line is taken from Ref. \cite{BMV}).
\end{table*}

\end{document}